\documentclass[journal=ancac3,manuscript=article,layout=twocolumn]{achemso}
\setkeys{acs}{doi=true}

\pdfoutput=1
\usepackage[version=3]{mhchem} 
\usepackage[T1]{fontenc}       
\usepackage{xcolor}            
\usepackage{hyperref}
\usepackage{doi}

\author{Sarah Friedensen}
\affiliation{Department of Physics and Astronomy, University of Pennsylvania, Philadelphia, Pennsylvania 19104, USA}

\author{William M. Parkin}
\affiliation{Department of Physics and Astronomy, University of Pennsylvania, Philadelphia, Pennsylvania 19104, USA}

\author{Jerome T. Mlack}
\affiliation{Department of Physics and Astronomy, University of Pennsylvania, Philadelphia, Pennsylvania 19104, USA}

\author{Marija Drndi\'{c}}
\affiliation{Department of Physics and Astronomy, University of Pennsylvania, Philadelphia, Pennsylvania 19104, USA}\email{drndic@sas.upenn.edu}

\title{TEM Nanosculpting of Topological Insulator \ce{Bi2Se3}}

\abbreviations{AFM, \ce{Bi2Se3}, FIB, FFT, HRTEM, PMMA, SAED, TEM}

\begin{document}

%
%
%
%
%

\begin{abstract}
We present a process for sculpting \ce{Bi2Se3} nanoflakes into application-relevant geometries using a high resolution transmission electron microscope. This process takes several minutes to sculpt small areas and can be used to cut the \ce{Bi2Se3} into wires and rings, to thin areas of the \ce{Bi2Se3}, and to drill circular holes and lines. We determined that this method allows for sub 10-nm features and results in clean edges along the drilled regions. Using in-situ high-resolution imaging, selected area diffraction, and atomic force microscopy, we found that this lithography process preserves the crystal structure of \ce{Bi2Se3}. TEM sculpting is more precise and potentially results in cleaner edges than does ion-beam modification; therefore, the promise of this method for thermoelectric and topological devices calls for further study into the transport properties of such structures.
\end{abstract}

\section{Introduction}
The material bismuth selenide (\ce{Bi2Se3}) has been well-studied for both its excellent thermoelectric properties\cite{GLSun_2015} and for its topologically-protected conductive surface states\cite{Shi_2015}. These states are most relevant when the ratio of surface conduction to bulk conduction is high. This occurs with large surface-to-volume ratio, so nanostructures of \ce{Bi2Se3} are of particular interest \cite{GLSun_2015}. In addition, theoretical proposals have suggested that specific nanostructure geometries may be able to enhance the material properties of \ce{Bi2Se3} in useful ways. For example, patterning an antidot lattice into \ce{Bi2Se3} may greatly improve its thermoelectric figure of merit\cite{Tretiakov_2011,Tretiakov_2012}, and pairing \ce{Bi2Se3} wires and rings with a superconductor may allow for the existence and control of Majorana fermions and other topological phenomena\cite{Ilan_2014}. Tuning the thickness of nanostructures may also permit control over transport phenomena \cite{Kang_2016,Sacksteder_2015}. Implementing these ideas is challenging, however, because most growth methods yield nanostructures in a great variety of sizes and shapes during the same process. This limits the ability to build devices to exact specifications and furthermore limits future scalability of geometry-dependent devices. One way to surmount these issues is to instead employ top-down fabrication methods and etch nanostructures into the precise shapes and sizes demanded by theory. Many such etching processes exist, and one method that allows for sub-10 nm feature sizes is transmission electron microscope (TEM) nanosculpting \cite{PDas_2016,Fischbein_2008}. TEM nanosculpting is a milling/dry etching method similar to focused ion beam (FIB) milling, but instead of using ions to ablate the material, it employs the high-energy electrons from a TEM beam. This process has been tested in several other materials and has been shown to allow control at the nanometer scale with minimal damage to the material structure \cite{Fischbein_2008,PDas_2016}. While other methods of etching have been applied to \ce{Bi2Se3} and the related compound \ce{Bi2Te3}\cite{ASharma_2016,PASharma_2014}, and while the effects of higher-energy irradiation on \ce{Bi2Se3} have been investigated\cite{Saji_2005}, the authors are unaware of research towards the use of TEM sculpting on \ce{Bi2Se3}.

In this paper we show results from the TEM sculpting of \ce{Bi2Se3} nanoflakes. By focusing the electron beam of the TEM on specific regions, Bi and Se were ablated from nanostructures, which allowed us to define a T-junction, a ring, a wire, a constriction, and an antidot array. Additionally, scanning the beam repeatedly over an area, we created a thinned region. We demonstrate that this process can be used to design sub-100 nm features in \ce{Bi2Se3} while leaving the nearby crystal structure largely intact.

\section{Results and discussion}

\begin{figure*}[ht]
\centering
\includegraphics{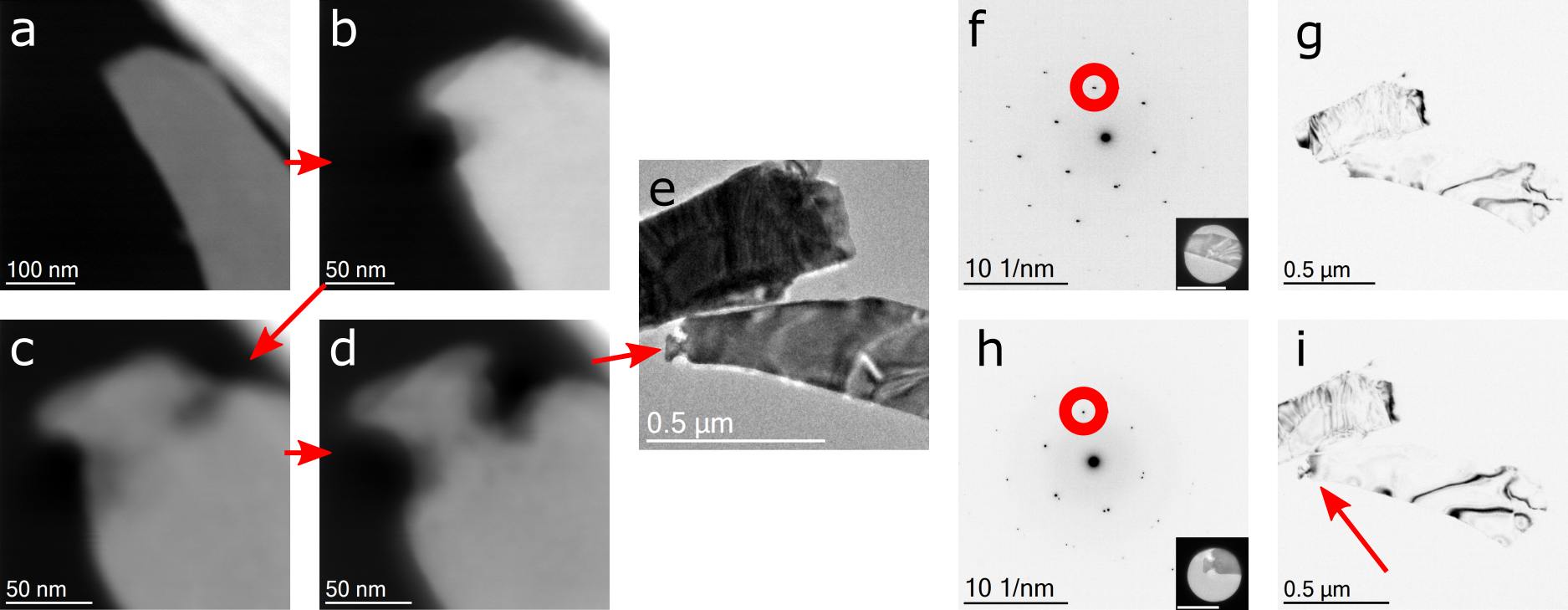}
\caption{\label{T_Junction_cut}
STEM images of a \ce{Bi2Se3} flake as it is cut to create a small T-junction. (a)-(d) Images taken sequentially during STEM drilling. In these and other dark-field images, the \ce{Bi2Se3} nanostructures appear in white on a black SiN$_x$ background. (e) TEM image of final structure. (f) SAED of initial structure at drilling site with selected area inset, aligned to [001] zone axis. Red-circled spot was used for dark-field imaging. Scale bar on inset is 0.5 $\mu$m. (g) Dark-field image of initial nanostructure (contrast inverted). (h) SAED of final structure with selected area inset. Red-circled spot again used for dark-field imaging. Scale bar on inset is 200 nm. (i) Dark-field image of final nanostructure (contrast inverted). Red arrow indicates location of T-junction.
}
\end{figure*}

\hyperref[T_Junction_cut]{Figure \ref*{T_Junction_cut}} documents the process of TEM sculpting a \ce{Bi2Se3} nanoflake using the "spot drilling" method.\footnote{This flake was sculpted in the "face-down" configuration. Sample orientations in the TEM and sculpting techniques are detailed in the ``Methods'' section} In this technique, the scanning probe focuses on single, user-defined regions of the sample for localized ablation. The dark-field STEM images of \hyperref[T_Junction_cut]{Fig. \ref*{T_Junction_cut}(a)-(d)} show the flake at various stages of the drilling process. The end result, shown in \hyperref[T_Junction_cut]{Fig. \ref*{T_Junction_cut}(e)}, is a T-junction of approximate length 80 nm, width 20 nm, and junction thickness 30 nm. \hyperref[T_Junction_cut]{Figs. \ref*{T_Junction_cut}(f) and \ref*{T_Junction_cut}(g)} show selected-area diffraction and dark-field imaging of the flake before modification. The contrast-inverted dark-field image in \hyperref[T_Junction_cut]{Fig. \ref*{T_Junction_cut}(g)} was taken using the circled diffraction spot. Similarly, \hyperref[T_Junction_cut]{Figs. \ref*{T_Junction_cut}(h) and \ref*{T_Junction_cut}(i)} present the selected area diffraction and inverted dark-field image of the flake after sculpting the T-junction. In \hyperref[T_Junction_cut]{Figs. \ref*{T_Junction_cut}(f) and \ref*{T_Junction_cut}(h)} , the flake is aligned along the [001] zone axis, and the measured lattice spacings of the \{110\} planes are 0.204 nm and 0.206 nm, respectively. These values agree with the known lattice spacing of 0.21 nm for \{110\} planes of \ce{Bi2Se3}\cite{Mlack_2013}. Additionally, splitting appeared in the lower three spots of the pattern in \hyperref[T_Junction_cut]{\ref*{T_Junction_cut}(h)}, which may indicate the lattice planes shifted across a dislocation. However, the darker band stretching from the T-junction into the body of the flake, shown in \hyperref[T_Junction_cut]{Fig. \ref*{T_Junction_cut}(i)}, indicates continuity along the selected lattice vector. Supplementary Figure S1 provides a second example of drilling a T-junction with images taken at 6-minute intervals during the drilling process as well as SAED patterns of this T-junction.

\begin{figure*}[ht]
	\centering
	\includegraphics{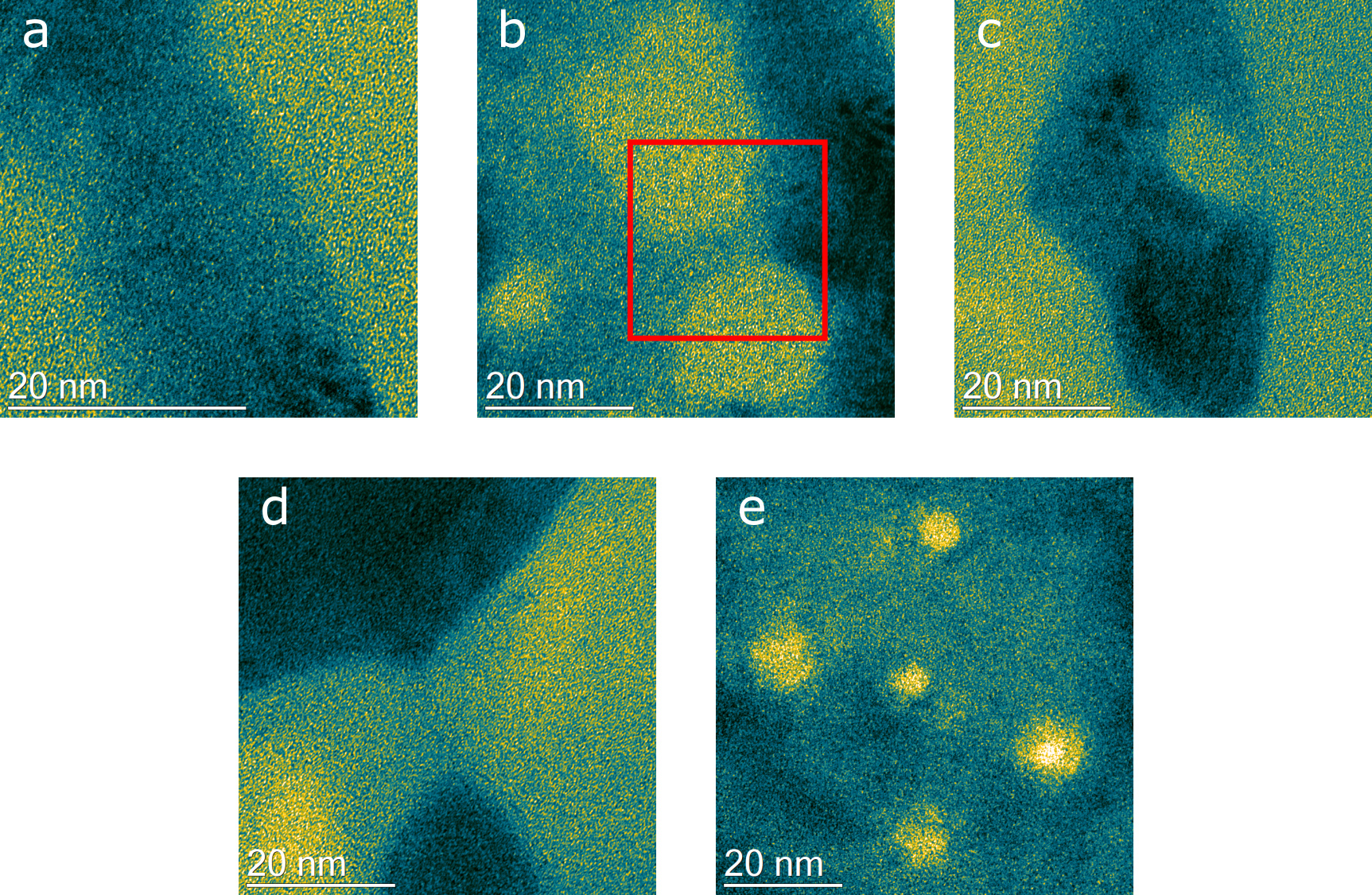}
	\caption{\label{Other_Sculptures_High_Res}
		False-color HRTEM of nanosculpted structures on a SiN$_x$ substrate. Blue indicates regions of lower transmission (typically thicker areas, where the \ce{Bi2Se3} is), and yellow indicates regions of higher transmission. (a) A 15 nm-wide wire (b) a thinner, shorter wire, boxed in red (c) a ring (d) a constriction (e) an antidot array. A grayscale image is available in Supplementary Figure S2, and the regions subject to STEM drilling are highlighted in Supplementary Figure S3.}
\end{figure*}

In addition to the T-junction, we sculpted structures including a long wire, a short wire, an annulus, a constriction, and a small antidot array using the same ``spot drilling'' process. \hyperref[Other_Sculptures_High_Res]{Fig. \ref*{Other_Sculptures_High_Res}} provides false-color high-resolution images of these structures (grayscale images provided in Supplementary Figure S2), and \hyperref[Other_Sculptures_FFT]{Fig. \ref*{Other_Sculptures_FFT}} shows their FFTs. Supplementary Figure S3 highlights in green the regions subject to irradiation with the probe. The FFTs shown in \hyperref[Other_Sculptures_FFT]{Fig. \ref*{Other_Sculptures_FFT}} suggest preservation of the intra-layer hexagonal crystal structure after sculpting. In general, the six-fold rotation symmetry survives, and measured lattice plane spacings are consistent with the \{110\} planes of bismuth selenide. In \hyperref[Other_Sculptures_FFT]{Fig. \ref*{Other_Sculptures_FFT}(c)} and \hyperref[Other_Sculptures_FFT]{Fig. \ref*{Other_Sculptures_FFT}(e)} the intensities of bismuth selenide spots in the FFT images varies, and in \hyperref[Other_Sculptures_FFT]{Fig. \ref*{Other_Sculptures_FFT}(d)} a new pair of spots appeared corresponding to a lattice spacing of 0.360 nm, which, similar to previous work with focused-ion-beam sculpting of \ce{Bi2Se3}, could correspond to a bismuth oxide or bismuth metal \cite{Friedensen_2017}. The formation of this phase is less prevalent than in the FIB sculpting, however. Enlarged copies of \hyperref[Other_Sculptures_FFT]{Figs. \ref*{Other_Sculptures_FFT}(b)}, \hyperref[Other_Sculptures_FFT]{\ref*{Other_Sculptures_FFT}(c)}, and \hyperref[Other_Sculptures_FFT]{\ref*{Other_Sculptures_FFT}(e)} are presented in Supplementary Figure S4.

Generally, the edges of the sculpted regions appear free of redeposited or recrystallized material, and lattice planes can be observed at the edges as well as within the structures. This implies that face-down TEM sculpting is potentially ``cleaner'' and more precise than ion-beam modification \cite{Friedensen_2017}. FFTs of selected areas from the ring structure and antidot array are available in Supplementary Figure S5; these confirm the presence of \ce{Bi2Se3} lattice planes within nanometers of the drilled regions. SAED images of the sculpted areas, which corroborate the measurements performed on the HRTEM images, are provided in Supplementary Figure S6.

\hyperref[Thinned_Region]{Fig. \ref*{Thinned_Region}(a)} presents an image of flake thinned by rastering the scanning probe over an area, with an image of the flake before thinning inset. Thinning was performed in two locations on the flake: a large area in the lower-right corner (green box) was subject to moderate thinning, and a smaller area nearby (yellow box) was subject to more severe thinning that in one area completely ablated the \ce{Bi2Se3}. Spot drilling was also performed along the line circled in blue. \hyperref[Thinned_Region]{Fig. \ref*{Thinned_Region}(b)} shows the thinned regions at higher magnification, with the red-boxed area selected for HRTEM analysis. Shown in \hyperref[Thinned_Region]{Fig. \ref*{Thinned_Region}(c)} is the FFT of that region, with the direct image inset. The average lattice plane spacing is 0.248 nm. \hyperref[Thinned_Region]{Fig. \ref*{Thinned_Region}(d)} shows a higher-magnification image of the spot-drilled line in the lower left corner of the flake. Lastly, \hyperref[Thinned_Region]{Fig. \ref*{Thinned_Region}(e)} shows an AFM image of the flake, which indicates both a decrease in height of about 20 nm and an increase in surface roughness (Rq) from 1.4 nm to 3.5 nm in and around the thinned regions.\footnote{The depth profile for the yellow-boxed regions is limited by the horizontal resolution of the AFM.}  On the other hand, the area around the line subject to spot drilling shows no roughness increase or surface damage in the surrounding areas, consistent with highly localized damage. The AFM of the green-boxed region in \hyperref[Thinned_Region]{Fig. \ref*{Thinned_Region}(c)} also implies that less severe thinning may be used to introduce surface disorder, which could have implications for controlling  transport via trapping and slowing surface conduction \cite{Sacksteder_2015}.

\begin{figure*}[ht]
	\centering
	\includegraphics{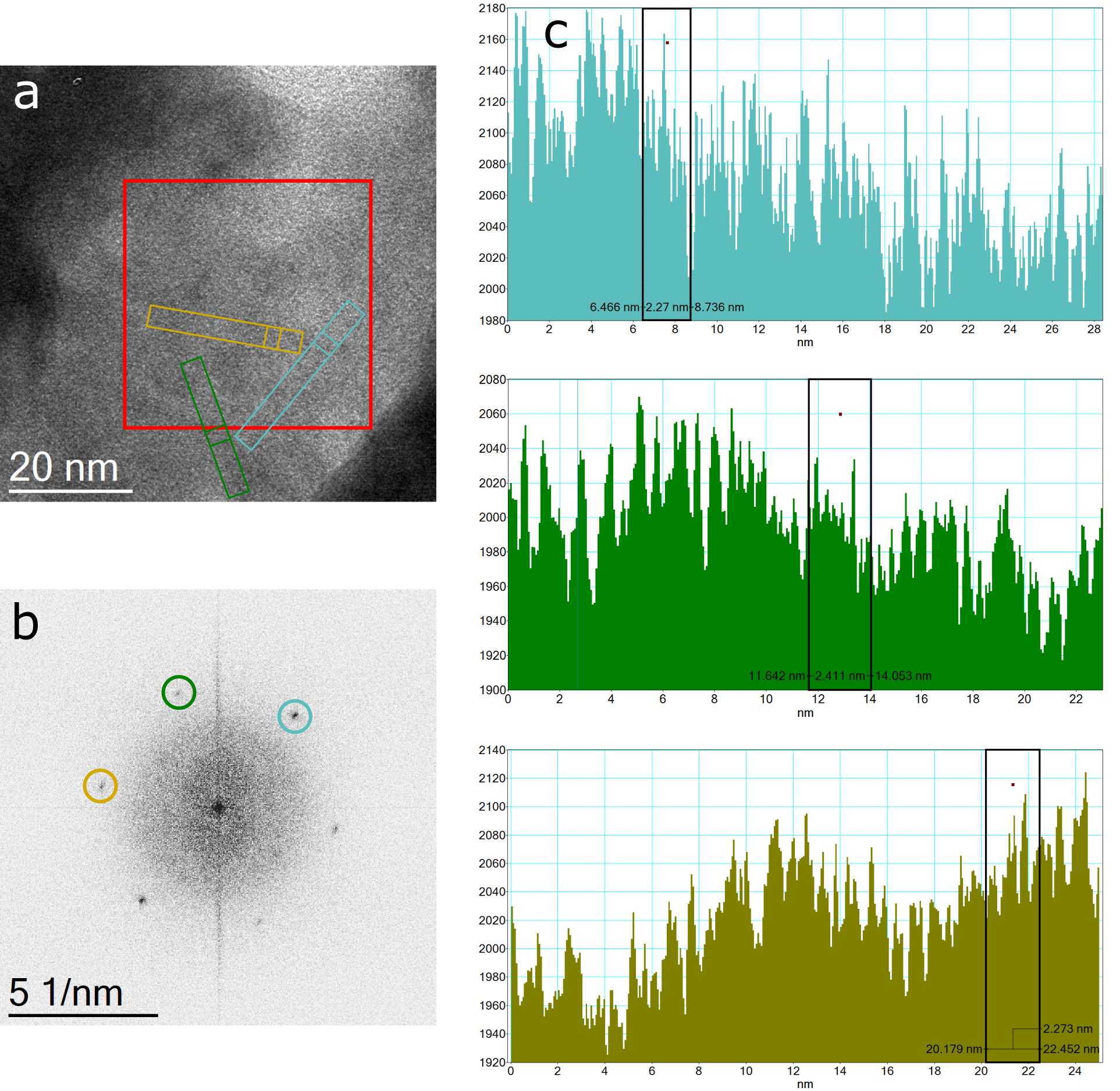}
	\caption{\label{Thinned_HR_Profile}
		Additional analysis of highly-thinned region. (a) HRTEM image of highly-thinned region. (b) FFT of red-boxed region in (a), which shows preservation of hexagonal symmetry even after extensive exposure to the beam. (c) Profile plots of (a), taken in the blue-, green-, and gold-boxed regions. The black boxes on the profiles enclose ten lattice fringes and show an average lattice spacing of 0.207 nm, up to software resolution. This is consistent with \ce{Bi2Se3}. The profiles were averaged over a 50-pixel width.}
\end{figure*}

\hyperref[Thinned_HR_Profile]{Fig. \ref*{Thinned_HR_Profile}(a)} shows an HRTEM image of the severely thinned area. The region boxed in red is one of the thinnest areas of the sample, and its FFT (\hyperref[Thinned_HR_Profile]{Fig. \ref*{Thinned_HR_Profile}(b)}) displays the hexagonal symmetry of \ce{Bi2Se3}. The teal, green, and tan boxes indicate 50 pixel-wide regions oriented along the \{110\} lattice vectors averaged for the profiles shown in \hyperref[Thinned_HR_Profile]{Fig. \ref*{Thinned_HR_Profile}(c)}, with like colors corresponding. The profiles run normal to the direction of the lattice planes. The black boxes in each profile measure the distance across ten lattice fringes---averaging these yields a lattice spacing of 0.232 nm. These images demonstrate that even though roughness increases in thinned regions, and even though these areas have experienced heavy electron beam exposure, the material itself retains its crystal structure. A larger AFM image is available as Supplementary Figure S7, and diffraction data in Supplementary Figure S8(a) and S8(b) supported these observations. 

\begin{figure*}[ht]
	\centering
	\includegraphics{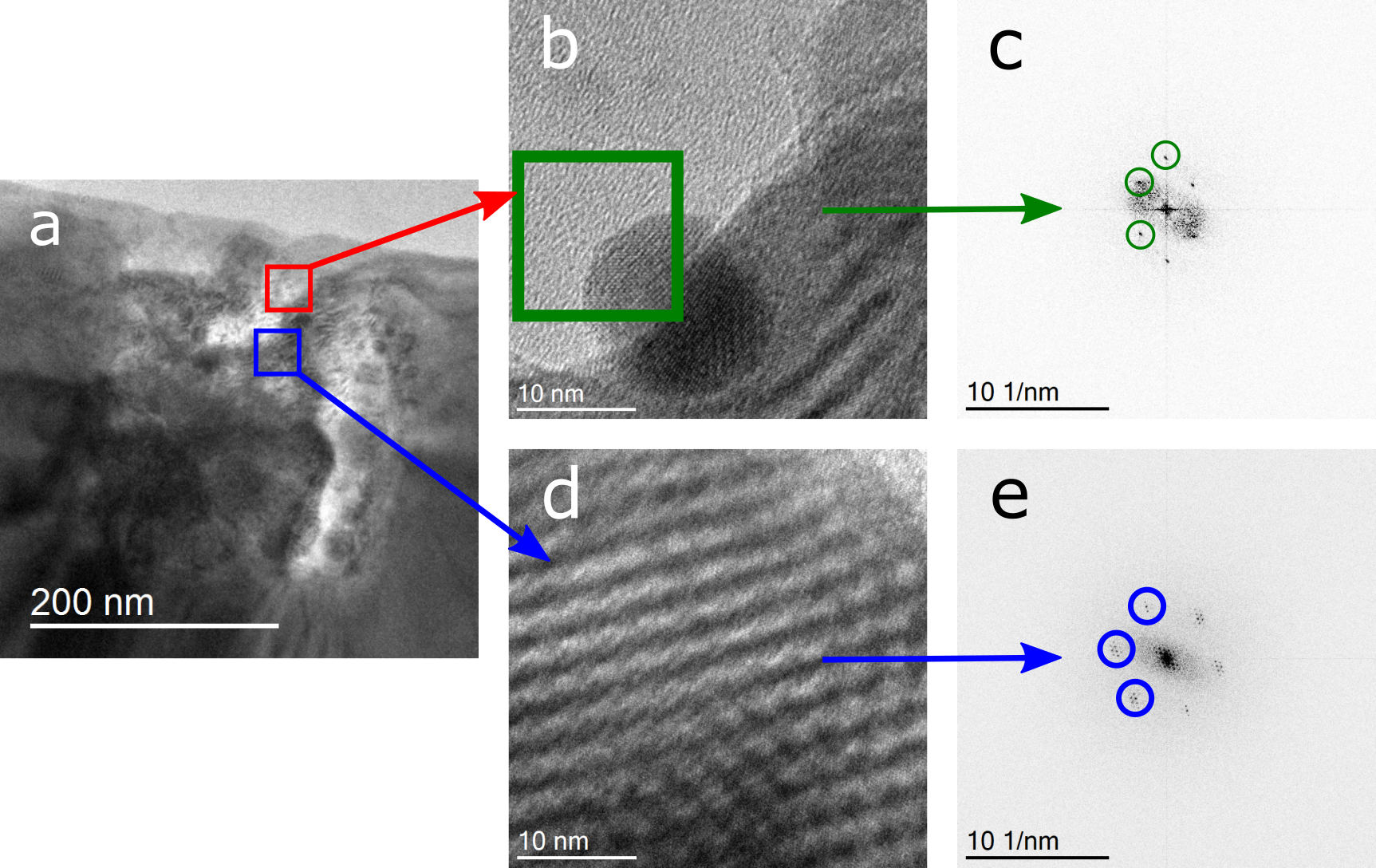}
	\caption{\label{WPS}
		Examples of structures formed from material redeposition during nanothinning with sample facing the electron source. (a) Image of the face-up thinned area. The red box encompasses the area shown in (b), and blue box encompasses that shown in (d). (b) A dot of recrystallized material formed during face-up thinning. Its resilience is apparent from its position over a hole drilled entirely through the surrounding \ce{Bi2Se3}. (c) FFT of the box outlined in (b). Clockwise from bottom left, the measured lattice plane spacings are 0.30 nm, 0.32 nm, and 0.27 nm. (d) Image of Moir\'{e} fringes formed under irradiation. (e) FFT of (d), showing superlattice.}
\end{figure*}

It is important for the sample to face downwards relative to the beam direction during sculpting---if the sample instead faces the source, ablated material redeposits on the substrate and forms crystalline structures that are not \ce{Bi2Se3} and are highly resistant to  further electron irradiation. These structures form during both spot drilling and large-area thinning and appear at random. Similar recrystallization has been observed during TEM sculpting of gold and platinum electrodes \cite{HZandbergen_2005}, though we did not observe elongation of the structures along the beam. \hyperref[WPS]{Fig. \ref*{WPS}(a)} shows the results of thinning with beam current $4.8 \cdot 10^9 \mathrm{A}/\mathrm{m}^2$ and the sample in a face-up configuration. While the exact nature of the redeposited structures was not determinable in this study, average lattice plane spacings of 0.30 nm and 0.32 nm measured in these regions could correspond to recrystallized bismuth or monoclinic bismuth oxide\cite{Friedensen_2017}. Given that the structures form under vacuum, we find that recrystallized bismuth is the more likely candidate. \hyperref[WPS]{Figs. \ref*{WPS}(b) and \ref*{WPS}(c)} show an example of such a structure and its FFT. Diffraction data shown in Supplementary Figure S8(c) generally agrees with these measurements. Irradiation in the face-up configuration also leads to Moir\'{e} patterns in the exposed areas, as shown in \hyperref[WPS]{Figs. \ref*{WPS}(d) \ref*{WPS}(e)}. We observed some interaction of these structures with the TEM beam under imaging conditions, which we detail in Supplementary Figure S9.

\section{Conclusions}
We have demonstrated the ability to sculpt \ce{Bi2Se3} flakes sitting on silicon nitride membranes with irradiation from a transmission electron microscope. Furthermore, we determined that this method allows for sub 10-nm features and results in clean edges along the drilled regions. Thinning the sample and introducing surface disorder across larger areas is possible by rastering the electron beam over an imaging window. The promise of this method for thermoelectric and topological devices calls for further study into the transport properties of such structures, particularly with respect to preservation and accessibility of electronic surface states.

\section{Methods}

We exfoliated nanostructures from bulk \ce{Bi2Se3} (99.999\% pure, Alfa Aesar) onto polymethyl methacrylate (PMMA). Using a micromanipulator and microscope method, we positioned and transferred selected flakes from the PMMA to a 100 nm-thick SiN$_x$ membrane\cite{JMlack_2017}. An acetone bath dissolved the PMMA, which completed the transfer of \ce{Bi2Se3} flakes to the membrane.

\begin{figure*}[ht]
	\centering
	\includegraphics{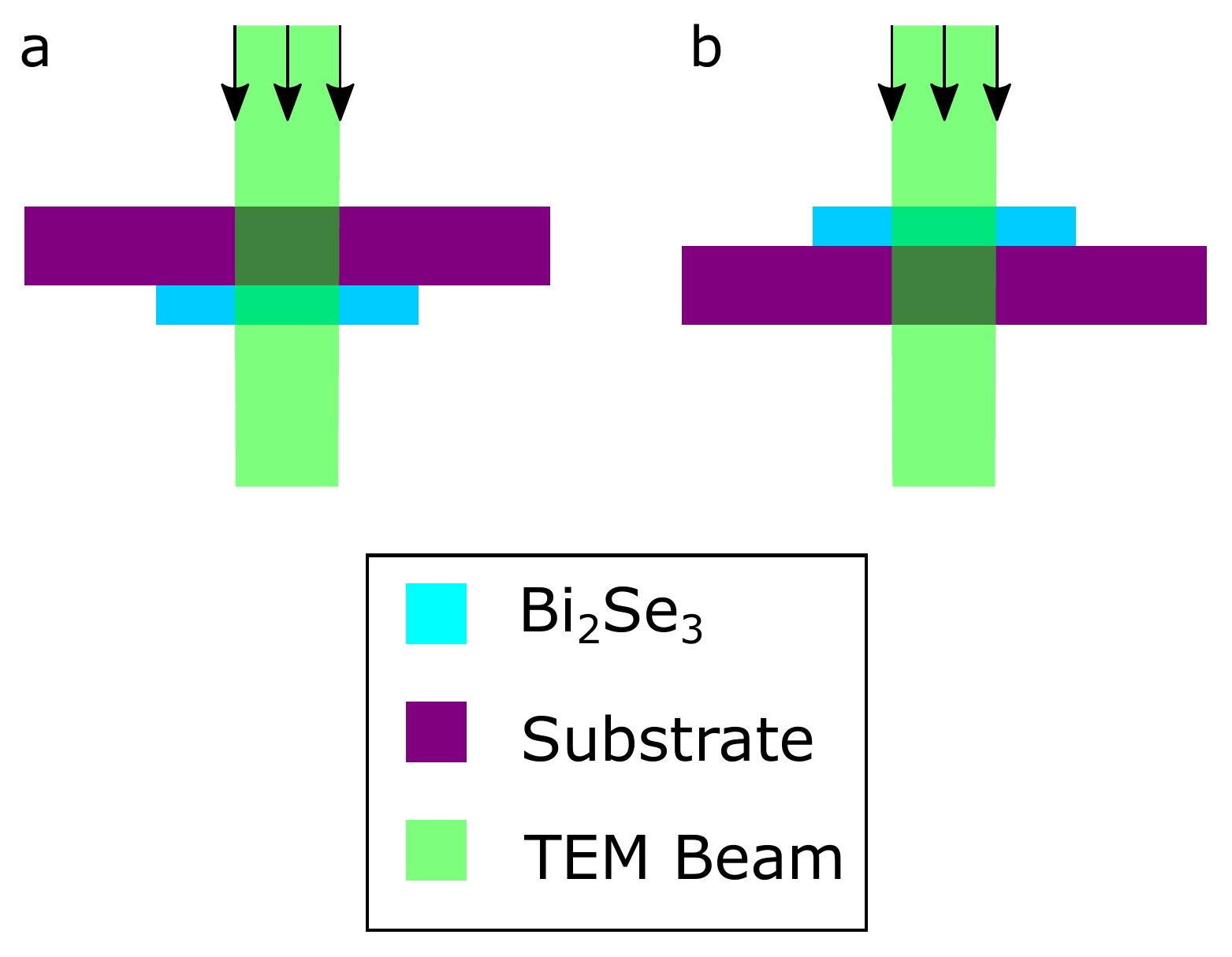}
	\caption{\label{beam_diagram}
		(a) Sample orientation relative to the beam in face-down configuration (b) in face-up configuration}
\end{figure*}

We sculpted and analyzed the using a JEOL 2010F TEM operating at 200 kV. The sample can be inserted into the TEM in two configurations: either the sample can face down, below the SiN$_x$ substrate and facing away from the beam source (\hyperref[beam_diagram]{Fig. \ref*{beam_diagram}(a)}), or face up, on top of the substrate and facing towards the beam source (\hyperref[beam_diagram]{Fig. \ref*{beam_diagram}(b)}). We found the face-down configuration superior for both high-resolution imaging and sculpting for these samples. During sculpting, we operated the TEM in scanning mode. In this mode, the electron probe is focused to a diameter of 2 nm or less, and the scan coils allow the probe to be positioned on the sample with nm precision. 

For sculpting using the ``spot drilling'' method, we set the beam current to $0.4 \cdot 10^9 \mathrm{A}/\mathrm{m}^2$, placed the beam in spot mode, and positioned the beam on the desired area of the sample. The probe impinged on this area of the sample for three to six minutes before we collected another scanned image to monitor progress and readjust the probe.

We thinned larger areas of the bismuth selenide using the following procedure, adapted from Rodr\'{i}guez-Manzo \latin{et al.}\cite{Rodriguez-Manzo_2015}. First, we increased the beam current from $0.4 \cdot 10^9 \mathrm{A}/\mathrm{m}^2$ to $4.8 \cdot 10^9 \mathrm{A}/\mathrm{m}^2$. While this modification negatively affects resolution, it is advisable for thinning to minimize process time. We then centered the write window on a chosen area, set the size and magnification (typically 84 nm $\cdot$ 84 nm) to define the thinning region, and increased the pixel time to 10 $\mu$s. We then imaged the sample continuously for intervals of fifteen minutes and continued the process until the thinning was sufficient for analysis.

\begin{figure*}[ht]
	\centering
	\includegraphics{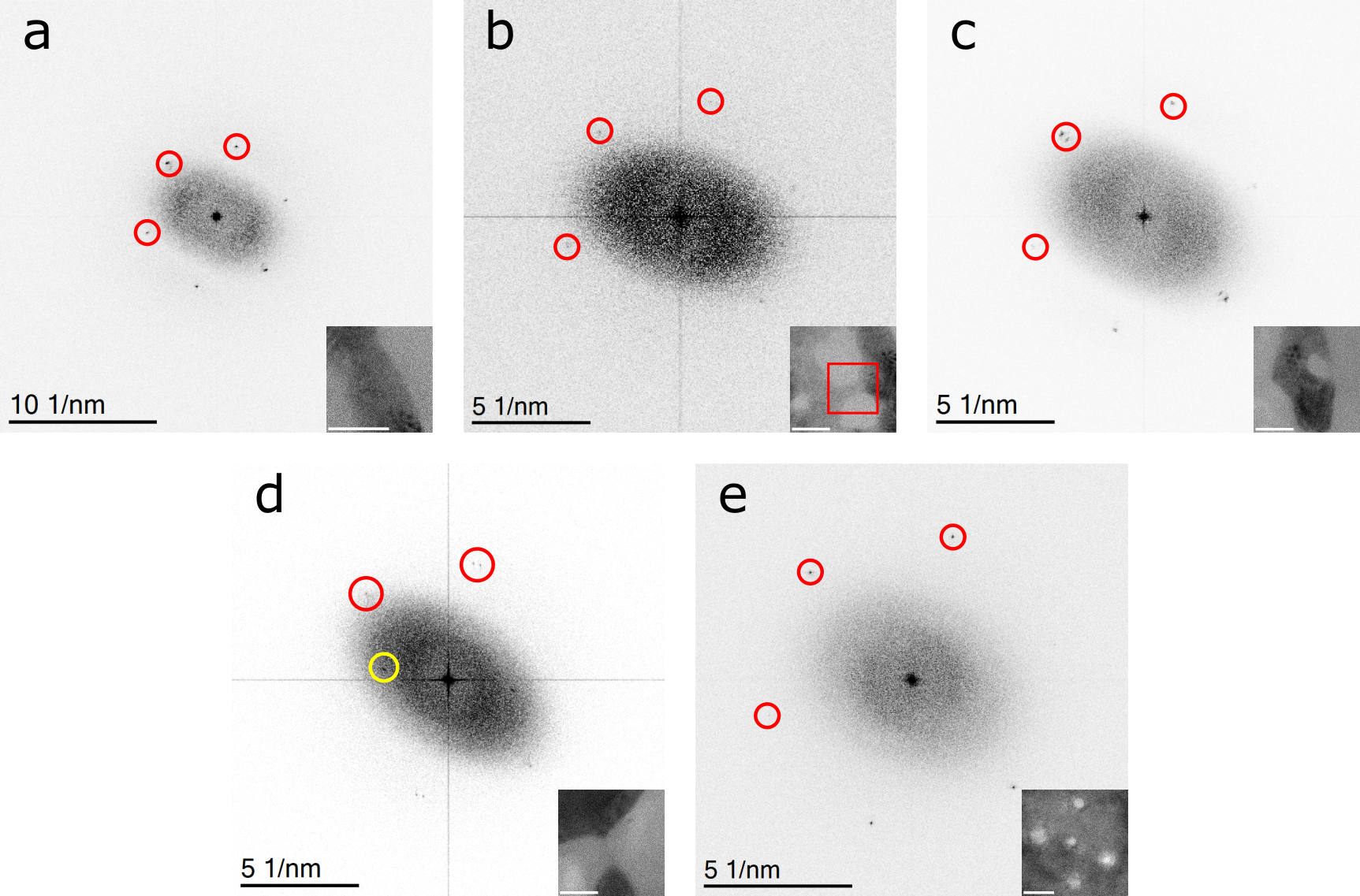}
	\caption{\label{Other_Sculptures_FFT} 
		FFTs of the images shown in \hyperref[Other_Sculptures_High_Res]{Figure \ref*{Other_Sculptures_High_Res}}. In all cases, spots corresponding to the (110) plane of \ce{Bi2Se3} appear. (a) The lattice plane spacing in the larger nanowire is measured to be 0.204 nm. (b) The measured lattice plane spacing in the smaller nanowire is 0.201 nm. This FFT corresponds to the red boxed area in \hyperref[Other_Sculptures_High_Res]{Figure \ref*{Other_Sculptures_High_Res}(b)}. (c) For the ring, the measured lattice plane spacing is 0.205 nm. One pair of spots shows twinning. (d) Lattice plane spacing is 0.199 nm. Additionally, one pair of spots has disappeared and a new pair of spots with spacing 0.360 nm appears. (e) Lattice plane spacing is 0.201 nm. All six spots are present, though one pair is very faint.}
\end{figure*}

\begin{figure*}[ht]
	\centering
	\includegraphics{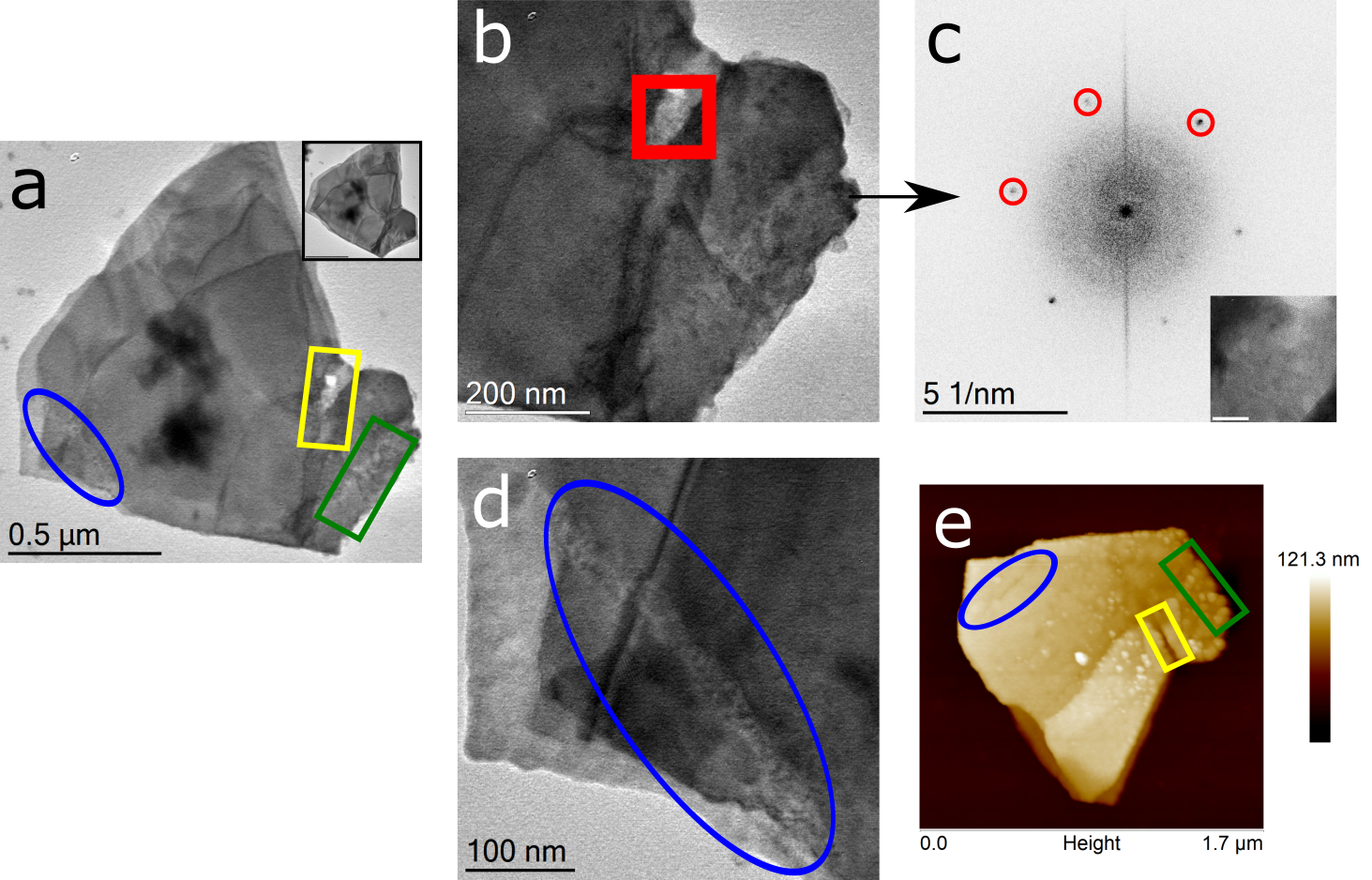}
	\caption{\label{Thinned_Region}
		Results of thinning \ce{Bi2Se3} by rastering over an area. (a) TEM image of flake after sculpting, with pre-sculpting image inset. Yellow and green boxes indicate regions subject to thinning; blue oval indicates additional spot-drilled region. (b) Higher-magnification image of the thinned regions of the flake. Red box indicates an area selected for HRTEM analysis. (c) FFT of red-boxed region near where exposure entirely ablated the \ce{Bi2Se3} (with direct image inset, scale bar 20 nm). Lattice spacing indicated by the FFT is 0.248 nm. (d) Higher-magnification of spot-drilled region, showing an extended divot. (e) AFM of thinned flake. Due to its orientation (face-down in TEM, face-up in AFM), the AFM image is mirrored relative to the TEM image.}
\end{figure*}

\begin{acknowledgement}
This material was supported in part by the National Science Foundation under EFRI 2-DARE 1542707 and through the NSF Graduate Research Fellowship Program under DGE-1321851 (S.F.). Any opinions, findings, and conclusions or recommendations expressed in this material are those of the authors and do not necessarily reflect the views of the National Science Foundation.

Use of University of Pennsylvania Nano/Bio Interface center instrumentation is acknowledged.

The authors sincerely thank Andrew Broeker for his suggestions regarding vibrational damping during TEM imaging.

\end{acknowledgement}

\begin{suppinfo}
	
	The following files are available free of charge.
	\begin{itemize}
			\item Supplementary\_Information.pdf: additional T-junction process images, grayscale structure gallery, structure gallery with highlighted drilled areas, selected-area diffraction and selected-area FFT images of sculpted and thinned structures, high-resolution images of behavior of the redeposited structures from face-up thinning during imaging, and larger copies of selected FFTs and AFM from the paper. 
		\end{itemize}

\end{suppinfo}

\doitext{}
\bibliography{Bi2Se3_Paper_Bibtex}

\providecommand{\latin}[1]{#1}
\makeatletter
\providecommand{\doi}
  {\begingroup\let\do\@makeother\dospecials
  \catcode`\{=1 \catcode`\}=2 \doi@aux}
\providecommand{\doi@aux}[1]{\endgroup\texttt{#1}}
\makeatother
\providecommand*\mcitethebibliography{\thebibliography}
\csname @ifundefined\endcsname{endmcitethebibliography}
  {\let\endmcitethebibliography\endthebibliography}{}
\begin{mcitethebibliography}{18}
\providecommand*\natexlab[1]{#1}
\providecommand*\mciteSetBstSublistMode[1]{}
\providecommand*\mciteSetBstMaxWidthForm[2]{}
\providecommand*\mciteBstWouldAddEndPuncttrue
  {\def\EndOfBibitem{\unskip.}}
\providecommand*\mciteBstWouldAddEndPunctfalse
  {\let\EndOfBibitem\relax}
\providecommand*\mciteSetBstMidEndSepPunct[3]{}
\providecommand*\mciteSetBstSublistLabelBeginEnd[3]{}
\providecommand*\EndOfBibitem{}
\mciteSetBstSublistMode{f}
\mciteSetBstMaxWidthForm{subitem}{(\alph{mcitesubitemcount})}
\mciteSetBstSublistLabelBeginEnd
  {\mcitemaxwidthsubitemform\space}
  {\relax}
  {\relax}

\bibitem[Sun \latin{et~al.}(2015)Sun, Li, Qin, Li, Zou, Xin, Ren, Zhang, Li,
  and Li]{GLSun_2015}
Sun,~G.~L.; Li,~L.~L.; Qin,~X.~Y.; Li,~D.; Zou,~T.~H.; Xin,~H.~X.; Ren,~B.~J.;
  Zhang,~J.; Li,~Y.~Y.; Li,~X.~J. Enhanced thermoelectric performance of
  nanostructured topological insulator Bi{$_{2}$}Se{$_{3}$}. \emph{Applied
  Physics Letters} \textbf{2015}, \emph{106}, \doi{10.1063/1.4907252}\relax
\mciteBstWouldAddEndPuncttrue
\mciteSetBstMidEndSepPunct{\mcitedefaultmidpunct}
{\mcitedefaultendpunct}{\mcitedefaultseppunct}\relax
\EndOfBibitem
\bibitem[Shi \latin{et~al.}(2015)Shi, Parker, Du, and Singh]{Shi_2015}
Shi,~H.; Parker,~D.; Du,~M.-H.; Singh,~D.~J. Connecting Thermoelectric
  Performance and Topological-Insulator Behavior:
  ${\mathrm{Bi}}_{2}{\mathrm{Te}}_{3}$ and
  ${\mathrm{Bi}}_{2}{\mathrm{Te}}_{2}\mathrm{Se}$ from First Principles.
  \emph{Phys. Rev. Applied} \textbf{2015}, \emph{3}, 014004, \doi{10.1103/PhysRevApplied.3.014004}\relax
\mciteBstWouldAddEndPuncttrue
\mciteSetBstMidEndSepPunct{\mcitedefaultmidpunct}
{\mcitedefaultendpunct}{\mcitedefaultseppunct}\relax
\EndOfBibitem
\bibitem[Tretiakov \latin{et~al.}(2011)Tretiakov, Abanov, and
  Sinova]{Tretiakov_2011}
Tretiakov,~O.~A.; Abanov,~A.; Sinova,~J. Holey topological thermoelectrics.
  \emph{Applied Physics Letters} \textbf{2011}, \emph{99}, 
  \doi{10.1063/1.3637055}\relax
\mciteBstWouldAddEndPuncttrue
\mciteSetBstMidEndSepPunct{\mcitedefaultmidpunct}
{\mcitedefaultendpunct}{\mcitedefaultseppunct}\relax
\EndOfBibitem
\bibitem[Tretiakov \latin{et~al.}(2012)Tretiakov, Abanov, and
  Sinova]{Tretiakov_2012}
Tretiakov,~O.~A.; Abanov,~A.; Sinova,~J. Thermoelectric efficiency of
  topological insulators in a magnetic field. \emph{Journal of Applied Physics}
  \textbf{2012}, \emph{111}, \doi{10.1063/1.3672847}\relax
\mciteBstWouldAddEndPuncttrue
\mciteSetBstMidEndSepPunct{\mcitedefaultmidpunct}
{\mcitedefaultendpunct}{\mcitedefaultseppunct}\relax
\EndOfBibitem
\bibitem[Ilan \latin{et~al.}(2014)Ilan, Bardarson, Sim, and Moore]{Ilan_2014}
Ilan,~R.; Bardarson,~J.~H.; Sim,~H.-S.; Moore,~J.~E. Detecting perfect
  transmission in Josephson junctions on the surface of three dimensional
  topological insulators. \emph{New Journal of Physics} \textbf{2014},
  \emph{16}, 053007, \doi{10.1088/1367-2630/16/5/053007}\relax
\mciteBstWouldAddEndPuncttrue
\mciteSetBstMidEndSepPunct{\mcitedefaultmidpunct}
{\mcitedefaultendpunct}{\mcitedefaultseppunct}\relax
\EndOfBibitem
\bibitem[Kang \latin{et~al.}(2016)Kang, Ha, Jung, Park, Song, Kim, and
  Hong]{Kang_2016}
Kang,~S.~M.; Ha,~S.-S.; Jung,~W.-G.; Park,~M.; Song,~H.-S.; Kim,~B.-J.;
  Hong,~J.-I. Two-dimensional nanoplates of Bi{$_{2}$}Te{$_{3}$} and
  Bi{$_{2}$}Se{$_{3}$} with reduced thermal stability. \emph{AIP Advances}
  \textbf{2016}, \emph{6}, \doi{10.1063/1.4942113}\relax
\mciteBstWouldAddEndPuncttrue
\mciteSetBstMidEndSepPunct{\mcitedefaultmidpunct}
{\mcitedefaultendpunct}{\mcitedefaultseppunct}\relax
\EndOfBibitem
\bibitem[Sacksteder \latin{et~al.}(2015)Sacksteder, Ohtsuki, and
  Kobayashi]{Sacksteder_2015}
Sacksteder,~V.; Ohtsuki,~T.; Kobayashi,~K. Modification and Control of
  Topological Insulator Surface States Using Surface Disorder. \emph{Phys. Rev.
  Applied} \textbf{2015}, \emph{3}, 064006, \doi{10.1103/PhysRevApplied.3.064006}\relax
\mciteBstWouldAddEndPuncttrue
\mciteSetBstMidEndSepPunct{\mcitedefaultmidpunct}
{\mcitedefaultendpunct}{\mcitedefaultseppunct}\relax
\EndOfBibitem
\bibitem[Masih~Das \latin{et~al.}(2016)Masih~Das, Danda, Cupo, Parkin, Liang,
  Kharche, Ling, Huang, Dresselhaus, Meunier, and Drndi{\'c}]{PDas_2016}
Masih~Das,~P.; Danda,~G.; Cupo,~A.; Parkin,~W.~M.; Liang,~L.; Kharche,~N.;
  Ling,~X.; Huang,~S.; Dresselhaus,~M.~S.; Meunier,~V. \latin{et~al.}
  Controlled Sculpture of Black Phosphorus Nanoribbons. \emph{ACS Nano}
  \textbf{2016}, \emph{10}, 5687--5695, \doi{10.1021/acsnano.6b02435},
  PMID: 27192448\relax
\mciteBstWouldAddEndPuncttrue
\mciteSetBstMidEndSepPunct{\mcitedefaultmidpunct}
{\mcitedefaultendpunct}{\mcitedefaultseppunct}\relax
\EndOfBibitem
\bibitem[Fischbein and Drndi{\'c}(2008)Fischbein, and
  Drndi{\'c}]{Fischbein_2008}
Fischbein,~M.~D.; Drndi{\'c},~M. Electron beam nanosculpting of suspended
  graphene sheets. \emph{Applied Physics Letters} \textbf{2008}, \emph{93}, \doi{10.1063/1.2980518}\relax
\mciteBstWouldAddEndPuncttrue
\mciteSetBstMidEndSepPunct{\mcitedefaultmidpunct}
{\mcitedefaultendpunct}{\mcitedefaultseppunct}\relax
\EndOfBibitem
\bibitem[Sharma \latin{et~al.}(2016)Sharma, Bhattacharyya, Shrivastava,
  Senguttuvan, and Husale]{ASharma_2016}
Sharma,~A.; Bhattacharyya,~B.; Shrivastava,~A.~K.; Senguttuvan,~T.~D.;
  Husale,~S. High performance broadband photodetector using fabricated
  nanowires of bismuth selenide. \emph{Scientific Reports} \textbf{2016},
  \emph{6}, \doi{10.1038/srep19138}\relax
\mciteBstWouldAddEndPuncttrue
\mciteSetBstMidEndSepPunct{\mcitedefaultmidpunct}
{\mcitedefaultendpunct}{\mcitedefaultseppunct}\relax
\EndOfBibitem
\bibitem[Sharma \latin{et~al.}(2014)Sharma, Lima~Sharma, Hekmaty, Hattar,
  Stavila, Goeke, Erickson, Medlin, Brahlek, Koirala, and Oh]{PASharma_2014}
Sharma,~P.~A.; Lima~Sharma,~A.~L.; Hekmaty,~M.; Hattar,~K.; Stavila,~V.;
  Goeke,~R.; Erickson,~K.; Medlin,~D.~L.; Brahlek,~M.; Koirala,~N.
  \latin{et~al.}  Ion beam modification of topological insulator bismuth
  selenide. \emph{Applied Physics Letters} \textbf{2014}, \emph{105}, \doi{10.1063/1.4904936}\relax
\mciteBstWouldAddEndPuncttrue
\mciteSetBstMidEndSepPunct{\mcitedefaultmidpunct}
{\mcitedefaultendpunct}{\mcitedefaultseppunct}\relax
\EndOfBibitem
\bibitem[Saji \latin{et~al.}(2005)Saji, Ampili, Yang, Ku, and
  Elizabeth]{Saji_2005}
Saji,~A.; Ampili,~S.; Yang,~S.-H.; Ku,~K.~J.; Elizabeth,~M. Effects of doping,
  electron irradiation, H + and He + implantation on the thermoelectric
  properties of Bi{$_{2}$}Se{$_{3}$} single crystals. \emph{Journal of Physics:
  Condensed Matter} \textbf{2005}, \emph{17}, 2873, \doi{10.1088/0953-8984/17/19/005}\relax
\mciteBstWouldAddEndPuncttrue
\mciteSetBstMidEndSepPunct{\mcitedefaultmidpunct}
{\mcitedefaultendpunct}{\mcitedefaultseppunct}\relax
\EndOfBibitem
\bibitem[Mlack \latin{et~al.}(2013)Mlack, Rahman, Johns, Livi, and
  Markovi{\'c}]{Mlack_2013}
Mlack,~J.~T.; Rahman,~A.; Johns,~G.~L.; Livi,~K. J.~T.; Markovi{\'c},~N.
  Substrate-independent catalyst-free synthesis of high-purity Bi2Se3
  nanostructures. \emph{Applied Physics Letters} \textbf{2013}, \emph{102},
  193108, \doi{10.1063/1.4807121}\relax
\mciteBstWouldAddEndPuncttrue
\mciteSetBstMidEndSepPunct{\mcitedefaultmidpunct}
{\mcitedefaultendpunct}{\mcitedefaultseppunct}\relax
\EndOfBibitem
\bibitem[Friedensen \latin{et~al.}(2017)Friedensen, Mlack, and
  Drndi{\'c}]{Friedensen_2017}
Friedensen,~S.; Mlack,~J.~T.; Drndi{\'c},~M. Materials analysis and focused ion
  beam nanofabrication of topological insulator Bi2Se3. \emph{Scientific
  Reports} \textbf{2017}, \emph{7}, 13466, \doi{10.1038/s41598-017-13863-6}\relax
\mciteBstWouldAddEndPuncttrue
\mciteSetBstMidEndSepPunct{\mcitedefaultmidpunct}
{\mcitedefaultendpunct}{\mcitedefaultseppunct}\relax
\EndOfBibitem
\bibitem[Zandbergen \latin{et~al.}(2005)Zandbergen, van Duuren, Alkemade,
  Lientschnig, Vasquez, Dekker, and Tichelaar]{HZandbergen_2005}
Zandbergen,~H.~W.; van Duuren,~R. J. H.~A.; Alkemade,~P. F.~A.;
  Lientschnig,~G.; Vasquez,~O.; Dekker,~C.; Tichelaar,~F.~D. Sculpting
  Nanoelectrodes with a Transmission Electron Beam for Electrical and
  Geometrical Characterization of Nanoparticles. \emph{Nano Letters}
  \textbf{2005}, \emph{5}, 549--553, \doi{10.1021/nl050106y}, PMID:
  15755112\relax
\mciteBstWouldAddEndPuncttrue
\mciteSetBstMidEndSepPunct{\mcitedefaultmidpunct}
{\mcitedefaultendpunct}{\mcitedefaultseppunct}\relax
\EndOfBibitem
\bibitem[Mlack \latin{et~al.}(2017)Mlack, Masih~Das, Danda, Chou, Naylor, Lin,
  Perea~L{\'o}pez, Zhang, Terrones, Johnson, and Drndi{\'c}]{JMlack_2017}
Mlack,~J.~T.; Masih~Das,~P.; Danda,~G.; Chou,~Y.-C.; Naylor,~C.~H.; Lin,~Z.;
  Perea~L{\'o}pez,~N.; Zhang,~T.; Terrones,~M.; Johnson,~A. T.~C.
  \latin{et~al.}  Transfer of monolayer TMD {WS$_{2}$} and Raman study of
  substrate effects. \emph{Scientific Reports} \textbf{2017}, \emph{7},   \doi{10.1038/srep43037}\relax
\mciteBstWouldAddEndPuncttrue
\mciteSetBstMidEndSepPunct{\mcitedefaultmidpunct}
{\mcitedefaultendpunct}{\mcitedefaultseppunct}\relax
\EndOfBibitem
\bibitem[Rodr{\'i}guez-Manzo \latin{et~al.}(2015)Rodr{\'i}guez-Manzo, Puster,
  Nicola{\"i}, Meunier, and Drndi{\'c}]{Rodriguez-Manzo_2015}
Rodr{\'i}guez-Manzo,~J.~A.; Puster,~M.; Nicola{\"i},~A.; Meunier,~V.;
  Drndi{\'c},~M. DNA Translocation in Nanometer Thick Silicon Nanopores.
  \emph{ACS Nano} \textbf{2015}, \emph{9}, 6555--6564, \doi{10.1021/acsnano.5b02531}, PMID: 26035079\relax
\mciteBstWouldAddEndPuncttrue
\mciteSetBstMidEndSepPunct{\mcitedefaultmidpunct}
{\mcitedefaultendpunct}{\mcitedefaultseppunct}\relax
\EndOfBibitem
\end{mcitethebibliography}

\end{document}